# Enceladus and Titan:
# Emerging Worlds of the Solar System

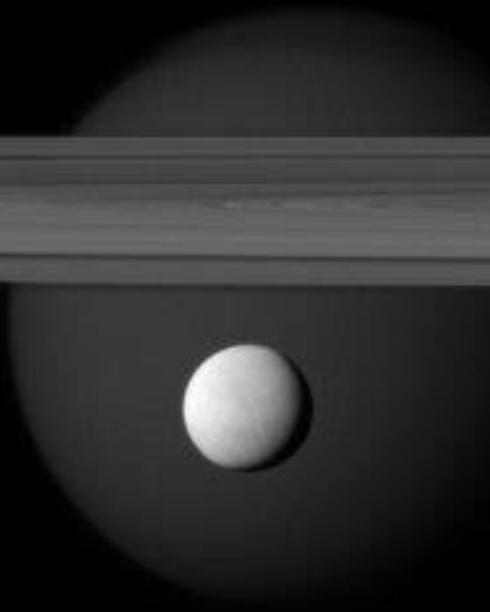

White paper submitted in response to Voyage 2050 long-term plan in the ESA Science Programme

Contact Scientist: Ali H. Sulaiman

Department of Physics and Astronomy
203 Van Allen Hall
The University of Iowa
Iowa City, IA 52242, USA
+1-319-335-0438
ali-sulaiman@uiowa.edu

Image of Saturn's rings, Titan and Enceladus taken by the camera onboard Cassini-Huygens
Credit: NASA/ESA

## Executive Summary

Some of the major discoveries of the recent *Cassini-Huygens* mission have put Titan and Enceladus firmly on the Solar System map. The mission has revolutionised our view of Solar System satellites, arguably matching their scientific importance with that of their planet. While Cassini-Huygens has made big surprises in revealing Titan's organically rich environment and Enceladus' cryovolcanism, the mission's success naturally leads us to further probe these findings. **We advocate the acknowledgement of Titan and Enceladus science as highly relevant to ESA's long-term roadmap,** as logical follow-on to Cassini-Huygens. In this white paper, we will outline important science questions regarding these satellites and identify the pertinent science themes we recommend ESA cover during the Voyage 2050 planning cycle. Addressing these science themes would make major advancements to the present knowledge we have about the Solar System, its formation, evolution and likelihood that other habitable environments exist outside the Earth's biosphere.

## Table of Contents

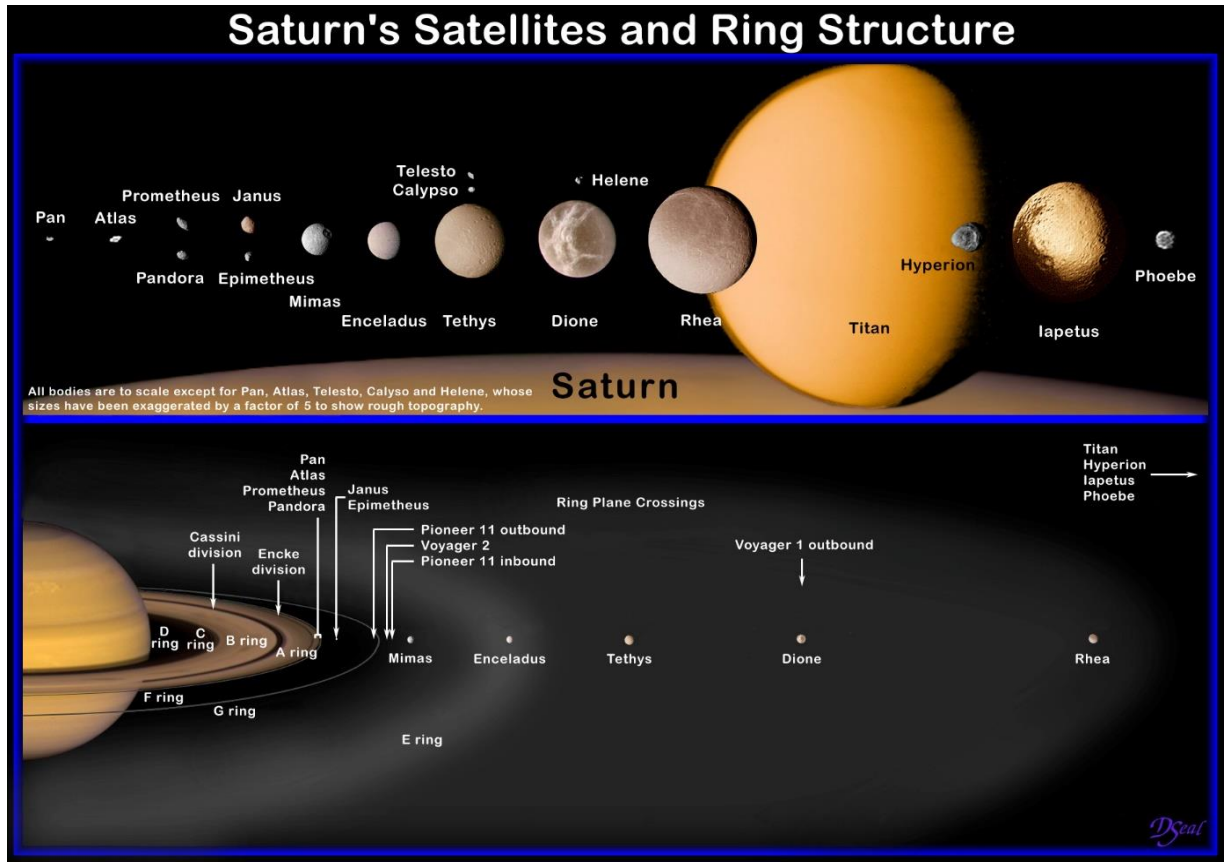

Figure 1 – Locations and scales of major satellites in Saturn's system. Credit: NASA/JPL

# 1. Introduction

## Why Explore Titan?

From *Voyager 1*'s glimpse of a hazy atmosphere to the successful entry and landing of the *Huygens* probe, Saturn's largest moon, Titan, remains to be an enigmatic Solar System body. Arguably the closest resembling Solar-System body to the Earth, Titan boasts a diverse landscape of lakes and rivers that are kept 'flowing' by the methane cycle - a striking parallel with the water cycle on Earth. Moreover, its thick, hazy atmosphere is sustained by a whole host of chemical processes that create complex organic compounds. For these reasons, we advocate Titan exploration as one of ESA's science priorities in the pursuit of emerging worlds in our Solar System and its potential to inform us about exotic exoplanetary systems.

## Why Explore Enceladus?

Enceladus is another unique planetary body. It is a small active moon that hides a global ocean under its thick icy crust. In its south polar region, the ocean material escapes through cracks in the ice. The escaping material forms a large plume of salty water that is rich in organic chemical compounds. Such key chemicals, in concert with ongoing hydrothermal activities and a tidally heated interior, make Enceladus a prime location for the search of a habitable world beyond the Earth. Enceladus science is highly relevant to ESA's goals in the next planning cycle and we recommend the acknowledgement that exploring Enceladus can make major advancements, as well as provide a unique opportunity to answering outstanding questions on habitability and the workings of the Solar System.

## 2. Overarching Science Themes

The exploration of Titan and Enceladus will address science themes that are central to ESA's existing Cosmic Vision programme, particularly on **habitability** and **workings of the Solar System**. The remarkable discoveries revealed by *Cassini-Huygens*, led to the proposal of a Large-class mission in response to the Cosmic Vision call with the goal of exploring Titan and Enceladus (Coustenis et al., 2009). The proposal was accepted for further studies, however did not go further. Over the last decade, numerous NASA missions have been proposed to build on the successes of *Cassini-Huygens* and explore these emerging worlds. In June 2019, NASA selected *Dragonfly* as their next New Frontiers mission to advance the search for the building blocks of life on Titan.

We advocate for these overarching themes since they encompass some of humankind's biggest open questions and should therefore remain a priority in ESA's next planning cycle. Missions to Titan and Enceladus would not only be a natural and logical follow-on to the successes of *Cassini-Huygens*; it would provide optimal laboratories to test questions pertaining to these overarching themes, namely: (i) What are the conditions for the emergence of life? (ii) How does the Solar System work? (iii) How are planetary bodies formed and how do they evolve? In addition to bringing multi-disciplinary Solar System science, addressing these questions can enhance our knowledge of exoplanetary systems and therefore foster synergy between Solar System scientists and the rapidly growing community of exoplanetary scientists.

## 3. Science Themes for Titan

### Titan's Atmosphere

Titan is well-known for its extensive atmosphere (e.g. Niemann et al., 2005; Wahlund, 2005). Because of its composition and complex organic chemistry (Waite et al., 2007; Vuitton et al., 2014), Titan's atmosphere is thought to be similar to that of early Earth, making it an obvious choice for studies on the origin of life.

The first signs of significant chemical complexity in Titan's atmosphere came from *Voyager* images of the satellite, which was obscured by an orange haze with a blue outer layer at the top of the atmosphere. This hid the surface from the visible cameras, and led Sagan et al (1993) to suggest tholins at Titan. *Cassini*'s instruments were able to penetrate this with radar, infrared and visible imaging, and the *Huygens* probe descended through the atmosphere. From orbit, complex chemistry involving neutrals, cations and anions (Waite et al., 2007, Coates et al., 2007) was found.

Some of the remarkable new results from the *Cassini* mission included: the unexpected presence of heavy negatively charged molecular ions (up to 13,800 u/q) and dust/aerosol particles (e.g. Coates et al 2011; Desai et al., 2017) making up a global dusty ionosphere (Shebanits et al., 2017); the formation of a 'soup' of organic (pre-biotic) compounds, including contributions to Titan's signature orange haze, as shown in Figure 2 (e.g. Waite et al., 2007; Vuitton et al., 2009); the unexpected impact of the solar EUV on the un-Chapman-like ionosphere (Ågren et al., 2007).

Titan's atmospheric chemistry is initiated in the ionosphere (thermosphere) primarily by the solar EUV on the dayside and energetic particle influx on the nightside (e.g. Cravens et al., 2006; Shebanits et al., 2013). It should be noted that while remote sensing provides excellent overview of Titan's ionosphere (e.g. Kliore et al., 2011) , detailed studies require in-situ measurements, not in the least due to the influence of the heavy negative charge carriers – molecular ions and dust/aerosols.

It was postulated at Titan that the high mass anions would drift down through the atmosphere, as tholins, eventually reaching the surface causing the dunes and falling in the lakes. The observations showed the highest masses of anions at the lowest altitudes (Coates et al., 2009) with the density showing similar trends (Wellbrock et al., 2013). The first chemical models showed that the low mass anion species may be CN-, C3N- and C5N- (Vuitton et al., 2009). Chemical schemes are only beginning to provide theories as to how the larger species can be produced. Charging models may explain how species of ~100 u could be ionized and aggregated to form >10,000 u molecules (Lavvas et al. 2011; Lindgren et al. 2016), but only a few studies have looked at precise chemical routes for producing molecules >100 u. An example is the formation of Polycyclic Aromatic Hydrocarbons (PAH) and more complex tholins, which are prebiotic polymer molecules, from simple hydrocarbons and nitrogen available as aerosols in Titan's atmospheric haze. It is clear the whole picture of chemical chains connecting all species of molecules is unclear.

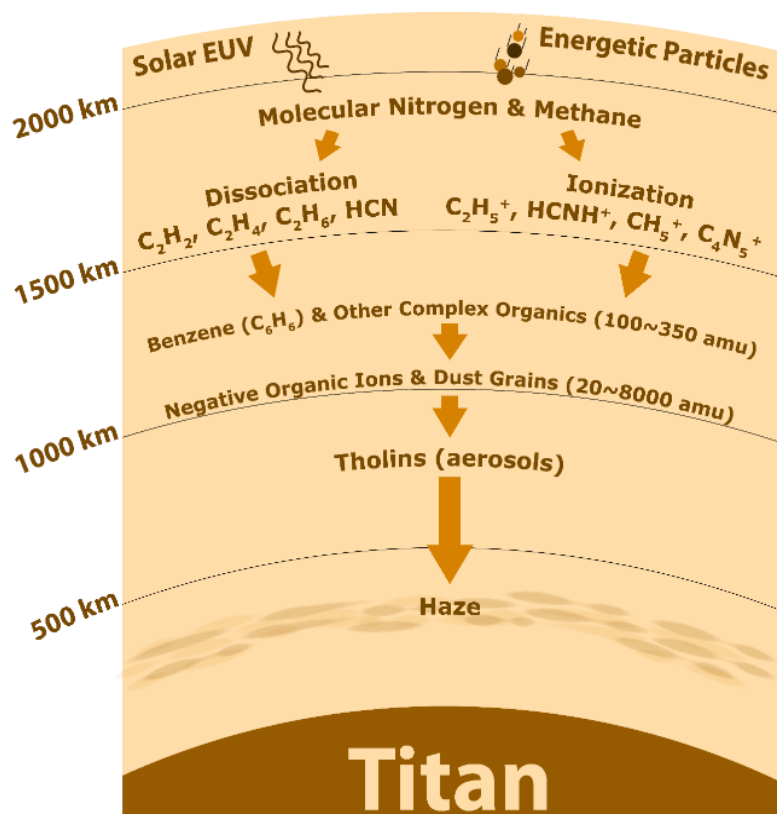

**Figure 2** – Schematic representation of Titan's complex atmospheric chemistry. Adapted from Waite et al., 2007.

Dusty plasma in Titan's ionosphere (e.g. Shebanits et al., 2016) (as well as Enceladus plume, e.g. Morooka et al., 2011) deserves special attention. Dusty plasma in space physics generally is a relatively new field, relevant for moon-produced plasma tori in Saturn and Jupiter systems, ionospheres of Earth (noctilucent clouds, e.g. Shukla, 2001) and Saturn (Morooka et al., 2019), cometary comas (e.g. Gombosi et al., 2015) and interstellar clouds (Sagan and Khare, 1979). For Titan's ionosphere, the dust in question may have different names (tholins, aerosols, dust grains, heavy negative ions) but generally refers to nm-sized grains or larger, with masses of more than a few hundred atomic mass units (e.g. Coates et al., 2007; Lavvas et al., 2013). The lack of consensus on the nomenclature in fact underlines the recency of the field.

Titan's dust grains form in the in-situ-accessible ionosphere and are indeed impossible to measure with the available remote sensing methods. Dusty plasma is also important for the energy budget as it increases ionospheric conductivities (Yaroshenko and Lühr, 2016). Aerosols/dust grains in general are also relevant to cloud formation (Anderson et al., 2018).

**Key measurements:** Ultraviolet, visible, infrared and millimeter/micrometer wave spectra will remotely constrain organic compounds in Titan's atmosphere and surface. High resolution mass/energy spectrometers and Langmuir probes will differentiate and constrain properties of neutrals, positive and negative ions, electrons and aerosols/dust grains. Radio occultations will resolve structure of the atmosphere.

*Summary*: **Titan's atmosphere is an excellent laboratory to study pre-biotic organic chemistry that directly ties into the question of the origin of life, and it is one of the available sites to study dusty plasma, a rapidly emerging field in the space community.**

> **Key scientific question:** What is the nature of atmospheric chemistry and cloud formation at Titan?

## Titan's Energy Budget

Titan is the largest moon in the Saturnian system and the only moon in the Solar System known to harbor a significant atmosphere. The moon lacks an intrinsic magnetic field and its radial distance from the planet at 20 Saturn radii (1 Saturn radius = 60,268 km) places it very close to the nominal distance of the subsolar outermost boundary of Saturn's magnetosphere, which is not static in response to variations in the solar wind dynamic pressure. This means Titan is at times inside the magnetosphere of Saturn, and at other times outside making it fully exposed to the solar wind. All of this adds to a very complex plasma interaction, where the moon can encounter not only the corotating plasma from the Saturnian magnetosphere, but also shocked solar wind (e.g. Wei et al., 2009) and even unperturbed solar wind (Bertucci et al., 2015).

*Atmospheric evolution and space weathering*

The surfaces and atmospheres of Saturnian moons are continuously irradiated with the magnetospheric plasma, solar photon, cosmic dust, and ring grains, all of which are responsible for long-term alteration of surface and atmospheric materials on geological timescales (Giga years), known as 'space weathering'. The space weathering accompanies dissociation and synthesis of molecules in the materials, followed by modification in surface and atmospheric spectra.

Titan represents the epitome of a non-stationary interaction and only through a detailed exploration of its environment (with a space weathering perspective), can the escape mechanisms and the amount of atmospheric loss to Saturn's magnetosphere and interplanetary space be appropriately addressed. The understanding of present-day escape conditions for Titan will contribute greatly to the elaboration of more realistic hypotheses about the evolution of Titan's atmosphere in the past and also the importance of atmospheric evolution in relation to habitability.

While Mars represents an evolved system embedded in the solar wind without a global magnetic field, where most of the atmosphere has been lost, Titan represents a system that, even without a global field, is protected by the Saturnian magnetosphere (most of the time) and the presence of a thick atmosphere, with a composition similar to that of the early Earth. Thus, understanding the relative contribution of the different escape mechanisms is important to further enhancing our current understanding of atmospheric evolution in the Solar System and the conditions for habitability both in our Solar System and in exoplanets and exomoons. During the *Cassini* era, several studies focused on understanding both the neutral escape (e.g. Tucker et al. 2009) and ion escape (e.g. Coates et al. 2012, Regoli et al. 2016). However, current datasets are neither sufficient nor instructive enough to separate the complex interaction due to the large variability of the upstream conditions.

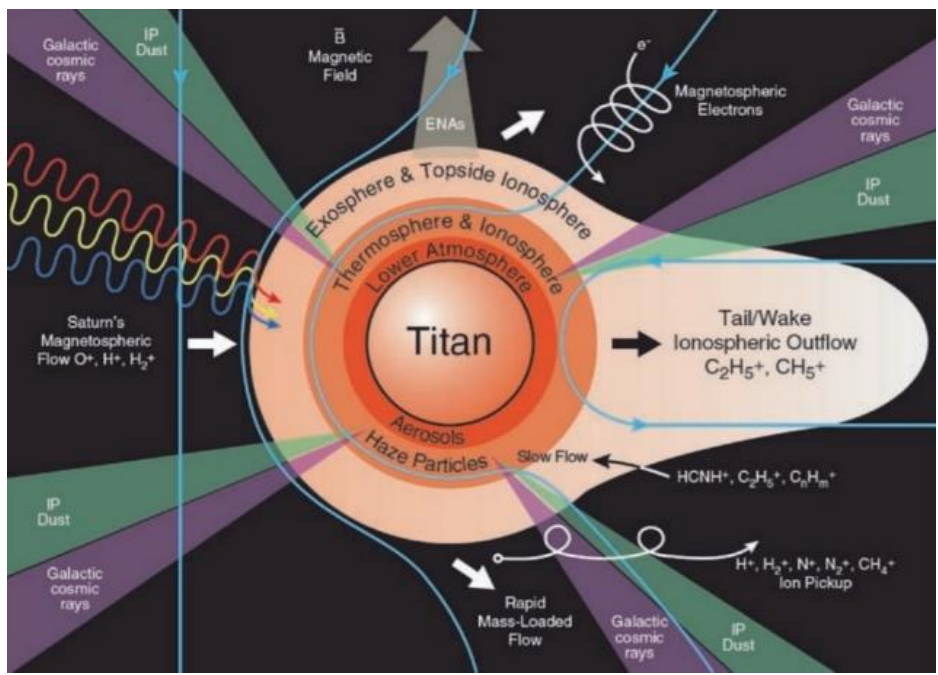

**Figure 3** – Energy and mass balance at Titan. Adapted from Sittler et al. (2009).

*An early Earth*

The atmosphere of Titan, composed mainly of $N_2$ (~94%) and $CH_4$ (~6%) has been likened to that of the primordial Earth, making this moon a natural environment to study processes that took place during the evolution of our own atmosphere. Modelling has shown Earth's early atmosphere may have been rich in hydrogen and methane (Tian et al., 2005). Moreover, key organic molecules in terrestrial prebiotic chemistry, such as hydrogen cyanide (HCN), cyanoacetylene ($HC_3N$) and cyanogen ($C_2N_2$), are formed at Titan (Teanby et al., 2006).

**Key measurements:** Plasma, neutral, and dust measurements will constrain influx and escape of material. Magnetic and electric field measurements will measure current systems that may arise and help constrain the energy budget of the system.

*Summary*: **Titan's atmosphere and surface are subjected to large variability in the ambient plasma and magnetic field by virtue of its orbit radius, continuous galactic cosmic rays, interplanetary dust, and solar photons. Altogether, these make the physical and chemical interactions of Titan and its environment inherently non-steady state at varying timescales.**

**Key scientific questions:** What is the response of Titan's atmosphere to extreme solar wind events, and how does it compare to quiescent events in the magnetosphere? What is the energy budget of Titan's atmosphere? Does Titan have an equilibrium state? How similar is Titan to an early Earth?

**Titan's Geology and Interior**

Titan has one of the most diverse surfaces across the Solar System, mainly due to its active methane cycle, which is somewhat analogous to the water cycle on Earth. Titan's complex surface has been modified by a variety of geological exogenic and/or endogenic processes. Some of the exogenic ones include impact cratering and aeolian/fluvial and/or lacustrine processes, while the endogenic ones include tectonism and potentially cryovolcanism (e.g. Elachi et al. 2005; Porco et al. 2005; Jaumann et al. 2009; Le Gall et al. 2010; Lopes et al. 2010; Wood et al. 2010).

Titan's surface investigation is complicated by its atmosphere. The thick and dense atmosphere of Titan can only be penetrated remotely in specific windows at near-infrared and radar wavelengths. Before 2004 and the entry of *Cassini* in the Saturnian system, bright and dark albedo Titan features were observed in near-infrared images taken by ground based telescopes and the Hubble Space Telescope (e.g. Coustenis et al. 2005). The 13 years of Titan investigation by *Cassini* and *Huygens*' landing on the surface in 2005 eventually revealed a remarkably Earth-like surface in terms of geomorphology, with dunes, highlands, dried and filled lakes, river channels and more. The *Cassini* investigation also unveiled Titan to be an organic-rich world (e.g. Janssen et al. 2016; Malaska et al. 2016; Hayes et al. 2018). Indeed, while Titan's crust is made of water ice, it is covered almost everywhere at the surface by a sedimentary organic layer of likely photochemical origin. The sediment materials are eroded, transported, and deposited from sources that are yet unclear, and organized to form landscapes that vary with latitude (dunes in the equatorial belts, plains at mid-latitude, labyrinthic terrains near the poles; Lopes et al. 2013; Solomonidou et al. 2018). The role of the methane cycle in the landscape distribution on Titan is yet to be understood. In addition, even if sedimentary processes seem to dominate Titan's surface, some features also suggest tectonism and cryovolcanism. The pursuit of the study of Titan's surface composition and its connection with the interior may unveil locations that are of importance to astrobiology and the search for life in the Solar System.

After the *Cassini* golden era and despite the great number of ground-breaking discoveries made by the 127 *Cassini* flybys of Titan, there are still open questions regarding the formation and evolution of the surface, its chemical composition, and the interactions between the surface, the interior, and the atmosphere. September 2017 marked the end of the *Cassini* mission; by then, ~65% of the surface had been imaged by the radar instrument with a spatial resolution in the range of 300m – 4km, and only ~20% of the surface had been captured by the Visual and Infrared Mapping Spectrometer (VIMS) with a resolution better than 10 km/pixel. It therefore remains terra incognita. The analysis of VIMS data provided significant results and insights on Titan's nature. However, the aforementioned resolution is not adequate for thorough and detailed investigation of the geology of a planetary body. In addition, the optimal use of surface data with *Cassini* instruments were made through the combinations of radar and VIMS data, which unfortunately were very limited due to the spacecraft's orbital constraints. In the future, such synergy between instruments would be of great value for a better understanding of Titan's geological history and evolution.

The *Cassini* and *Huygens* data, the multiple years of data analysis, laboratory studies, and theoretical and experimental modeling prompt the science goals for the future planetary missions to Titan. Below are a number of key features and processes at Titan.

*Impact craters*

The abundance and size distribution of impact craters usually provide insight into the relative age of planetary surfaces. Titan, compared to other Saturnian moons, displays a very limited number of impact craters on its surface, indicating a relatively young and active surface (e.g. Wood et al. 2010; Werinsky et al. 2019). Even though some studies have provided constraints (Radebaugh et al. 2008; Neish and Lorenz, 2012; Lopes et al. 2016), the surface age of Titan still remains uncertain (probably between 200 Ma and 1 Ga).

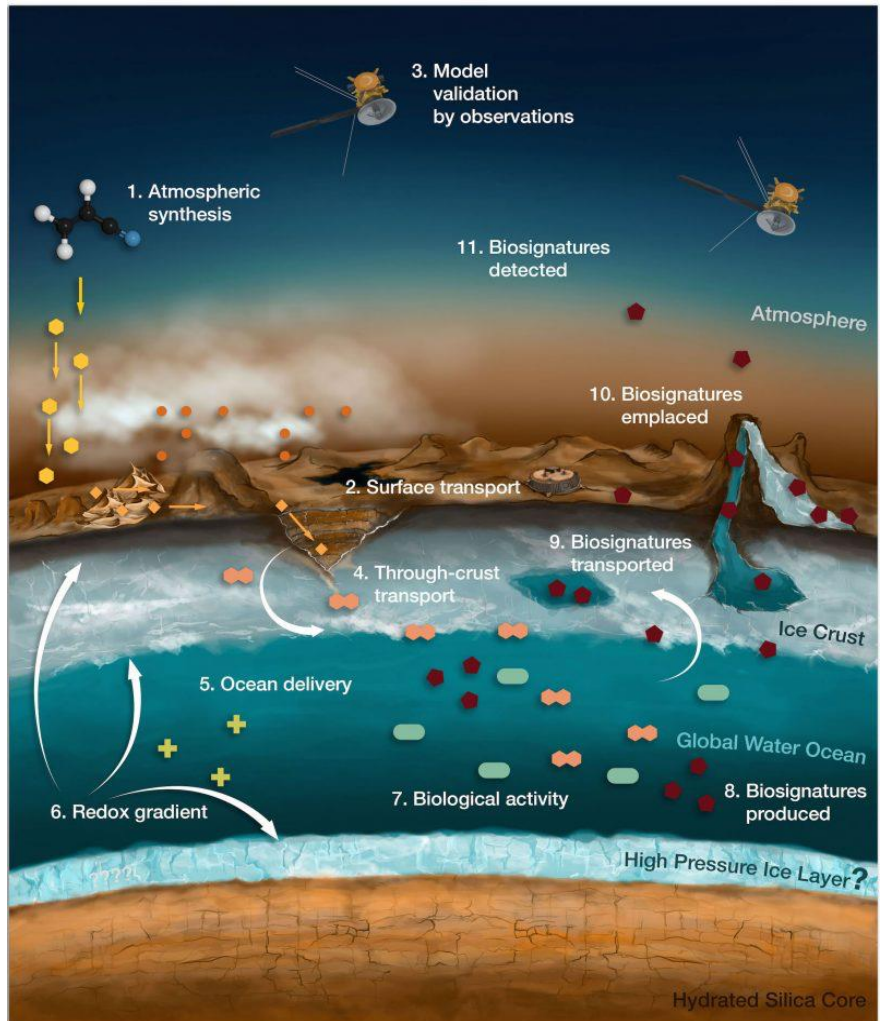

**Figure 4** – Diagram illustrating Titan's multiple layers from the atmosphere down to the interior and how biosignatures could also be transported from the subsurface ocean to the surface of Titan. (Image Credit: A. Karagiotas/T. Shalamberidze/NAI/JPL

*Winds*

One of the major geological processes on Titan's low latitudes is the aeolian (wind) activity. The dark, organic-rich terrains that dominate the equator are giant dune fields, which are hundreds of km long, a few km wide and about 50-150m in height. Strong winds occurring at the equinoxes seem to control their direction (Charnay et al. 2015), but local topography and ground humidity seem to play a role too. Further investigation of the dune morphometry and composition would help better understand Titan's meteorology and geology.

*Plains*

60% of the surface of Titan is covered by plains that appear uniform at the resolution of the *Cassini* radar (300 m at best). Future missions will have to unravel the mystery of these features, which likely hold important clues on the evolution of Titan's surface.

*Lakes*

Titan is the only other planetary body in the Solar System to possess bodies of liquid on its surface that are stable in time. *Cassini* observations of Titan have revealed three seas and ~650 polar lakes, 200 being empty and more than 300 filled or partially filled (e.g. Stofan et al. 2007; Hayes, 2016). Modeling suggests the liquid composition to be a mixture of methane and ethane with the contribution of dissolved nitrogen (e.g., Sagan and Dermott, 1982). However, *Cassini* data rather suggest a dominance of the methane (Mastrogiuseppe et al. 2014; 2018). It remains to explain the fate of ethane which is produced in abundance in the atmosphere by photochemistry. In addition, most of Titan's smaller lakes are characterised as sharp-edged depressions with raised rims and ramparts surrounding some of them (e.g. Birch et al. 2018; Solomonidou et al. 2019). The origin of both and, more generally, the formation mechanism of Titan's lakes remains unknown.

*Tectonism & Cryovolcanism*

Titan's tectonism and cryovolcanism is still, after *Cassini*'s exploration, under debate. Features on Titan such as linear dark features, ridges, mountains, and canyons (Porco et al. 2005; Radebaugh et al. 2007; Mitri et al. 2010; Lopes et al. 2010) suggest tectonism (Scheidegger, 2004) followed by exogenous processes. However, *Cassini* imagery is insufficient to understand the origin of the observed surface deformations (tides, crust cooling etc).

Another surface process that has not yet been identified, but is speculated, is cryovolcanism. Outgassing by cryovolcanism has been proposed as a possible replenishment mechanism for Titan's atmospheric methane (e.g. Lopes et al. 2007) and a number of plausible cryovolcanic landforms were proposed on the basis of their morphology (e.g. lobate flows in Sotra Patera) and/or because surface changes were observed (Solomonidou et al. 2016). However, there is no "smoking gun" for cryovolcanism on Titan and the idea of cryovolcanism as a possible shaping process remains controversial (e.g. Moore and Pappalardo, 2011). Future missions will shed light on the exchanges between the interior, the surface, and the atmosphere of Titan.

*Astrobiology*

Titan harbors a combination of complex hydrocarbons and organic molecules in addition to a water ocean beneath its ice shell and potential cryovolcanism. Liquid water can also exist at the surface for a limited time, e.g. after an impact or a cryoeruption. All these suggest conditions potentially suitable for life as we know it and future missions should investigate Titan's chemistry and search for biosignatures.

**Key measurements:** High spatial resolution radar and infrared spectrometer will map Titan's surface. Complete coverage will be achieved by an orbiter. Mass spectrometer will determine chemistry of Titan lakes. Sonar will determine depth of a Titan sea. Gravity measurements will characterise Titan's interior. High-resolution and high-sensitivity mass spectrometry will identify key molecules in search of biosignatures.

*Summary*: **Titan is arguably the most Earth-like Solar System body. Its methane cycle draws an analogy with Earth's water cycle. The abundance of organic material, water and energy source due to potential cryovolcanism qualifies the satellite as a prime candidate for a habitable world in the Solar System.**

> **Key scientific questions:** What are the characteristics of Titan's habitability and what potential biosignatures should we look for? What is the composition and distribution of materials on and beneath Titan's surface? What are the lakes made of? Is the interior active?

# Titan's Interaction with Saturn's Magnetosphere

Titan's orbit radius of 20 Saturn radii places it, most of the time, within Saturn's magnetosphere, and embedded in the rapidly rotating, magnetised plasma that flows at ~100 km s$^{-1}$, much faster than the Titan orbital speed of ~6 km s$^{-1}$. As the plasma flows past and around Titan, magnetic field lines that are 'frozen' into the moving plasma drape around the moon, thus forming downstream lobes in which the field generally points towards Titan in one lobe, and away in the other. The magnetic field configuration arising from this type of interaction depends on the field and plasma conditions upstream of Titan.

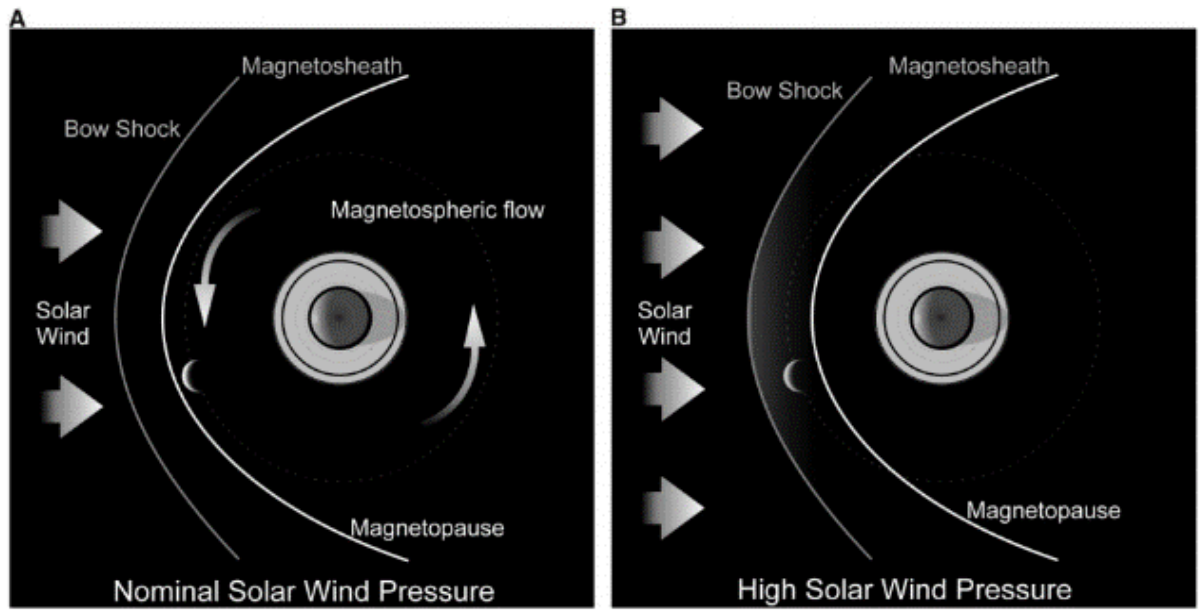

**Figure 5** – Sketch describing how Titan's plasma interaction depends on solar wind pressure. Under nominal solar wind conditions, Titan interacts with Saturn's rotating magnetosphere (**A**). When solar wind pressure is high, Titan exits Saturn's magnetosphere and is exposed to the solar wind (**B**). Figure adapted from Bertucci et al. (2008).

Before *Cassini*, the common perception was that the upstream field would be oriented north-south, the equatorial direction of Saturn's dipole field. The *Cassini* mission, however, has shown a very different picture. Saturn's disk-like plasma sheet continuously flaps up and down past Titan, with a period close to what we believe is the true rotation period of the planet – around 10.7 hours. This flapping does not come from any tilt in Saturn's magnetic equator (Saturn's internal field is almost perfectly aligned with the planet's rotational axis). Rather, it arises from a rotating wave-like pattern imposed on the sheet by rotating systems of electric currents in the magnetosphere, flowing on field lines extending ~10-15 Saturn radii. As the plasma sheet moves, the upstream field changes, being dominantly north-south when Titan is near the centre of the plasma sheet. These changes were characterised by Bertucci et al (2009), who surveyed *Cassini* magnetometer data during spacecraft flybys of Titan.

*Fossil magnetic field*

Later, Achilleos et al (2014) used a model of the plasma sheet (magnetodisk) to study one fly-by in detail. They found that the magnetospheric flux tubes that flow closest to Titan may carry with them the imprint of a very different kind of upstream field compared to the imprint carried by plasma in the far-Titan space. This is because the upstream field is continually changing. This change in magnetic 'imprint' could become even more pronounced if the boundary of Saturn's magnetosphere moves inward or outward past Titan. When this happens (albeit relatively rarely), Titan transitions between a magnetospheric and a solar wind regime, a process first discovered by Bertucci et al. (2008) in the T32 flyby of *Cassini*. At T32 closest approach (CA) to Titan, the magnetic field, remarkably, had a southward component. By contrast, the ambient solar wind magnetic field surrounding the CA interval had a northward component. The southward component near CA is consistent with the draped field that *would* have been seen, had Titan been continuously immersed in Saturn's magnetosphere throughout the encounter. But at CA, both Titan and *Cassini* were, unambiguously, outside the magnetosphere. In fact, Titan had been there for at least ~15 minutes, just after spending up to three hours in the magnetosphere. Hence, the field that was imprinted on Titan's ionosphere during its magnetospheric excursion survived there for at least 15 minutes. The time range of the (external) magnetic imprint on Titan was found to be ~15 minutes - ~3 hours, hence constraining the lifetime of the imprint *fossil field* at Titan, and raises the intriguing prospect of 'magnetic archeology', where close flybys of Titan could potentially reveal details of ambient fields to which Titan has been exposed up to about three hours in the past.

The interaction between Titan and the magnetosphere is bidirectional. While the incoming plasma and magnetic field defines many aspects of the interaction at the moon, the presence of the moon itself affects the local magnetosphere in ways that are not yet fully understood. The model of the interaction, as described by computer simulations (e.g. Simon et al. 2015) predicts the existence of a pair of extended Alfvén wings (standing Alfvén waves) that connect with the planet's ionosphere. However, these wings have not yet been detected, most probably due to the limited spatial coverage of the *Cassini* flybys.

### Internal magnetic field

A study of *Cassini* magnetometer data in the near-Titan environment by Wei et al. (2010) demonstrated that the upper limit of a putative permanent dipole moment would be ~0.78 nT $R_{Titan}^3$, not significantly different from zero. The lack of a magnetic dynamo inside Titan is consistent with the incompletely differentiated interior suggested by *Cassini* gravity measurements (Iess et al., 2010). The existence of an induced dipole moment, arising from the penetration of a slowly varying external field into a subsurface conducting region, remains an open question – the variation of the magnetospheric field on the ~29-year orbital timescale of Saturn, for example, may be a viable candidate for induction. Any induced dipole moment would be expected to change direction during Saturn's equinox, at which time average direction of the ambient magnetospheric field changes direction, due to the displacement of the mean current sheet from above to below the equator (for the equinox captured by *Cassini*). Wei et al. (2018) have recently reported finding a possible reversing induced field signature through comparing pre- and post-equinox field data from *Cassini*. This suggests that further characterisations or constraints on the induced dipole at Titan are needed for finally answering the question of whether an electrically conducting region, such as an ocean, exists beneath the surface.

**Key measurements:** Magnetometers will measure the fossil magnetic field of Titan and measure any induced magnetic field generated in Titan's interior. Electric field antennas, magnetometers and plasma spectrometers will measure any potential electrodynamic coupling between Titan and Saturn.

***Summary*: Titan's interaction with Saturn is both highly dynamic and bi-directional. It is unclear whether Titan has an internally generated magnetic field.**

> **Key scientific questions:** How exactly does Titan interact with Saturn? What is the power transfer between the two systems? Is there an induced magnetic field generated in Titan's interior that might be associated with a subsurface ocean or a weak dynamo?

## 4. Science Themes for Enceladus

### Enceladus' Plume

*Cassini* has also dramatically revolutionised our view of the small icy moon Enceladus, of ~500 km in diameter, orbiting at 4 Saturn radii. It was discovered that strong plumes emanate from warm 'tiger stripe' features on its south polar surface (Dougherty et al., 2006, Porco et al., 2006). These plumes of water vapour and ice grains are thought to be the long-suspected source of particles making up Saturn's E-ring, and also the dominant source for neutrals and plasma in Saturn's magnetosphere. In-situ observations here revealed primarily water vapour and trace amounts of hydrocarbon-based neutral gas (Waite et al., 2009), as well as water-group positive ions that slow, divert and even stagnate the magnetospheric flow (Tokar et al., 2009). Directly over the plume sources, charged nanograin populations have been observed that are related to the tiger stripes but dispersed in their motion by Saturn's magnetic field (Jones et al., 2009). Negative water group ions, possibly with additional species consistent with hydrocarbons, are also seen.

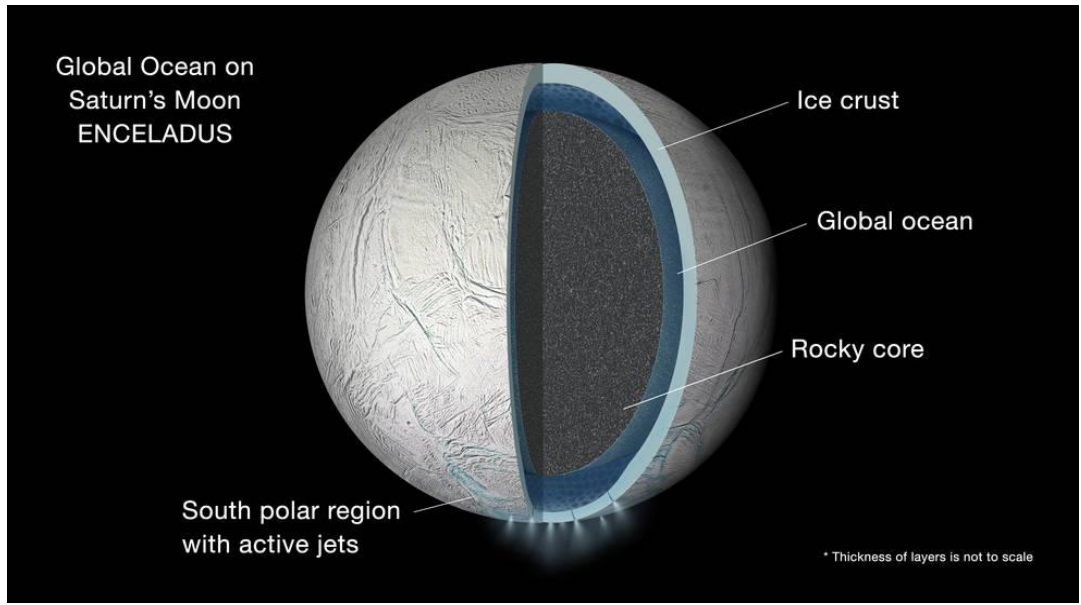

**Figure 6** – Illustration of the interior of Enceladus showing a global liquid water ocean between its rocky core and icy crust. Image credit: NASA/JPL-Caltech.

*Ongoing hydrothermal activities*

Repeated *Cassini* sampling of Enceladus' plume ejecta, simulations and laboratory experiments have concluded present-day hydrothermal activity at Enceladus may resemble what is seen in the deep oceans of Earth. The detection of sodium-salt-rich ice grains emitted from the plume suggests that the grains formed as frozen droplets from a liquid reservoir that has been in contact with rocks (Postberg et al., 2009). Gravity measurements suggest the presence of a subsurface sea at depths of 30-40 km and extending up to south latitudes of about 50° (Iess et al., 2014). These findings hint rock-water interactions in regions surrounding the core of Enceladus resulting in chemical 'footprints' being preserved in the liquid and subsequently transported upwards to the near-surface plume sources, where they are eventually ejected. Furthermore, the detection of nanometer-sized silica particles indicates ongoing high-temperature (>90 °C) hydrothermal reactions associated with global-scale geothermal activity. (Hsu et al., 2015; Sekine et al., 2015; Choblet et al., 2017).

*Tidal forces*

The brightness of the plume of Enceladus has been shown to depend on the orbital phase of Enceladus (Hedman et al., 2013). Since Enceladus' orbit is slightly elliptical, tidal stresses will act on the moon and the cracks in its south polar region will either be more or less open depending on the distance between Enceladus and Saturn. Hedman et al., 2013 showed that the plumes' brightness is several times greater when Enceladus is around apocenter (farthest away from Saturn) than when the moon is around pericenter (closest to Saturn). A change in plume brightness may be caused by a change in the size distribution of the grains and not solely to their total mass.

Several studies (e.g. Saur et al., 2008) have shown variability in the amount of water molecules ejected from the cracks in the icy crust of Enceladus. However, the correlation between the orbital phase of Enceladus and the modulation in the ejected material has been difficult to confirm (e.g. Hansen et al., 2015). The *Cassini* observations of the plume have raised many questions about the driving processes, the time modulations, the structure of the plume, and the dynamics of the ejected material.

The water that is ejected from the south polar region of Enceladus creates a neutral water torus around Saturn, along the orbit of the moon. Subsequently, transport, photoionization, and electron impact ionization of the neutral material creates a plasma disk at the same location. The disk has been suggested to vary with longitude (Gurnett et al., 2007), local time (Holmberg et al., 2014), season (Tseng et al., 2010) and the solar cycle (Holmberg et al., 2017). These asymmetries found in the plasma disk are still under investigation. A complicating factor to finding a clear modulation in the plasma disk is that the source of disk material, i.e., the Enceladus plume, is also varying.

*Astrobiology*

The Ion Neutral Mass Spectrometer (INMS) onboard *Cassini* sampled Enceladus' plume and found ammonia, along with various other organic compounds, deuterium and very probably $^{40}$Ar (Waite et al., 2009). Since ammonia acts as an anti-freeze, its presence is strong evidence for the existence of liquid water, given that the measured temperatures exceed 180 K near the fractures from which the jets emanate (Spencer et al., 2006). Temperatures were measured by *Cassini*'s Composite Infrared Spectrometer (CIRS), which detected 3-7 GW of thermal emission from the south polar troughs and confirming an internal heat source. This makes Enceladus the third known solid planetary body that is sufficiently geologically active for its internal heat to be detected by remote sensing – after Earth and Io.

INMS also detected molecular hydrogen in the plume (Waite et al, 2017). Ongoing hydrothermal reactions of rock containing reduced minerals and organic materials have been invoked as the most plausible source of this hydrogen. Waite et al. (2017) further postulated that the relatively high hydrogen abundance in the plume signals thermodynamic disequilibrium that favours the formation of methane from $CO_2$ in Enceladus' ocean. This state of disequilibrium is exploited by some forms of life (chemolithotrophs) as a source of chemical energy. $H_2$ metabolisms are used by some of the most phylogenetically ancient forms of life on Earth (Raymann et al., 2015), while on modern Earth, geochemical fuels such as $H_2$ support thriving ecosystems even in the absence of sunlight (Kelley et al., 2001).

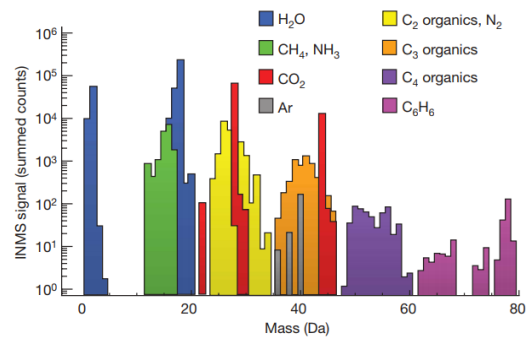

**Figure 7** – Composition of Enceladus' plume as measured by *Cassini*'s Ion and Neutral Mass Spectrometer (Waite et al., 2009).

That said, while *Cassini* has flown through and directly sampled Enceladus' plumes, it did so with instruments that were 20 years old and with very limited capabilities. The aforementioned discoveries cannot categorically confirm evidence of biological processes. More complex and highly sensitive analyses will recognise long chain molecules and amino acids that are uniquely interesting targets in the search for life. Even more complex analyses such as identifying the chirality (left-handed vs right-handed) of amino acids will be very instructive. Nearly all amino acids on Earth are left-handed since biological processes requires this basic consistency for proteins to fold. It is therefore expected that any protein-based life will "choose" a particular chirality, i.e. all left-handed or all right-handed, rather than an equal mixture of the two (Creamer et al., 2016).

*Dusty Plasmas*

In the plume, the exhaust from the south pole creates a different plasma regime, dusty plasma. When the ice particles from the fractures immerse into the ambient plasma, they acquire charges. The charge state of the particle varies depending on the surrounding plasma condition. In the dense plasma, as in Saturn's inner magnetosphere, the electrical potential of the dust becomes slightly negative (Kempf et al., 2006; Wahlund et al., 2009), and the grain charge number varies from single to several thousand (e.g., Horanyi et al., 1992; Yaroshenko et al., 2009). The charged grains from Enceladus in Saturn's magnetosphere are under the influences of both gravity as well as electromagnetic forces. When the number of charged grains is large and inter-grain distance is small compared to the plasma Debye length, charged dust particles participate in the collective behavior, i.e. dusty plasma in contrast to dust-laden plasma.

A number of Enceladus flybys by *Cassini* provided direct measurements of the dust and plasma in the plume regions. The charged grains have been directly confirmed by the Cosmic Dust Analyzer (CDA) (Kempf et al., 2008), the Radio Plasma Wave Science (RPWS) antenna (Kurth et al., 2006), and the plasma spectrometer (CAPS) as high energy charged particles (Hill et al., 2012). The total charge number of the grains has been inferred by the Langmuir probe and the magnetic field measurements (Morooka et al., 2011). Combining these measurements concluded that the typical size of the outgas from the south pole are nanometers to micrometers, however, the overall negative charge in the plume is carried by the nanometer to sub-micrometer grains (Dong et al., 2015), and they are in the dusty plasma regime (Morooka et al., 2011).

**Key measurements:** Modern chemical spectrometers will identify long chain molecules, such as essential amino acids required for biological processes, with the capability of discriminating between left-handed and right-handed chirality. Chemical spectrometers will also identify other important compounds, relative abundances and oxidation states that are key ingredients for biological processes. Infrared and ultraviolet spectrometers will monitor the plume gas, its activity and dust distributions. Dust analysers and plasma spectrometers (electron and ion) will measure a wide range of the size distribution of grains. Plasma spectrometers and Langmuir probes will measure electrical potential of the grains, as well as electron temperature and ion speeds. Gravity measurements will characterise Enceladus' interior.

*Summary:* **Enceladus has ongoing hydrothermal activity from its tidally heated interior. The plume emanates from the southern polar region and has been measured to contain water, volatiles and organic compounds. The plume originates from a subsurface salty ocean. Altogether, the knowns of (i) an accessible salty ocean, (ii) organic compounds, (iii) energy and (iv) hydrothermal activity make Enceladus a prime candidate to explore habitability outside the Earth's biosphere.**

**Key scientific questions:** Is life present in Enceladus now? What is the chemistry of its plume? How does the chemistry evolve over time? Does the chemistry contain signatures of biology?

## Enceladus' Interaction with Saturn's Magnetosphere

### Electric current system

The coupling between Enceladus and Saturn is in many aspects similar to the one observed near Jupiter's moon Io (Neubauer, 1980). It includes the existence of accelerated plasma and magnetically field-aligned electric currents, associated with Alfvén wings, producing *auroral* footprints on the Saturn's atmosphere. Indeed, two striking observations of the Enceladus auroral footprint have been reported by Pryor et al. 2011 and observations of accelerated electron beams associated with plasma wave emissions (Gurnett, et al. 2011) and field aligned electric currents (Engelhardt et al. 2015) have been also reported near the moon at the edge of the plume.

Since the plume is located near the south pole of Enceladus, a north-south asymmetry in Enceladus' plasma interaction and the Alfvén wing system is introduced. The south pole plume launches Alfvén waves, which are partially blocked by the solid body of Enceladus. This leads to hemispheric coupling currents along the Enceladus flux tube and associated discontinuities in the magnetic field (Saur et al. 2007, Simon et al. 2014). In addition to the spatial asymmetries of the interaction, the plasma interaction is also time-dependent due to the diurnal variability of Enceladus' plume activity (Saur et al. 2008, Hedman et al. 2013).

More recently when *Cassini* sampled Saturn's topside ionosphere, Sulaiman et al. (2018) reported observations of plasma processes and strong electric currents demonstrably linked to Enceladus. The detection of such phenomena when *Cassini* was so close to Saturn underlined the nonlocality and spatial extent of the ever-present coupling between Enceladus and Saturn, thus indicating these two bodies are in continuous energy exchange between each other. This magnitude of this energy, however, remains to be quantified.

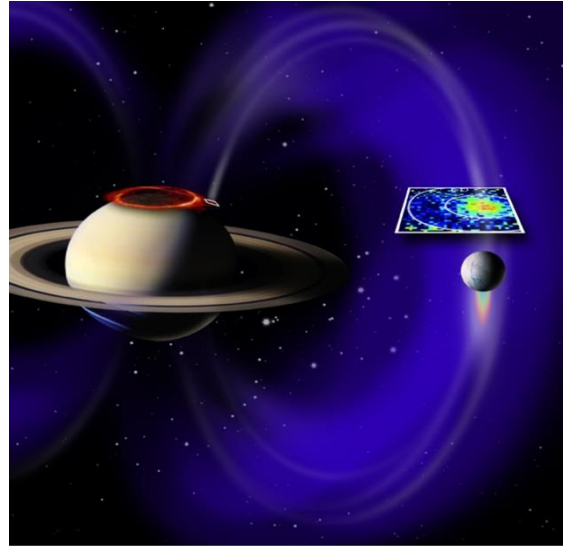

**Figure 8** – Artist's concept of the coupling between Enceladus and Saturn creating an auroral footprint on Saturn's atmosphere. Credit: K. Moscati and A. Rymer; JHU/APL

### Space weathering

As introduced earlier, prebiotic polymers are synthesized in Titan's atmosphere. Such a process could also happen on Enceladus' surface because there are source materials, e.g., ammonia and tholins, already suggested from surface reflectance at ultraviolet wavelengths (Zastrow et al., 2012).

Space weathering inhomogeneity potentially exists at Enceladus because it has an inhomogeneous magnetic field induced by electromagnetic interactions between Saturn's magnetosphere, Enceladus' surface, plume, and subsurface ocean (e.g., Jia et al., 2010).

If we successfully associate characteristics of the space weathering with the cumulative dose of irradiation, the duration of weathering can be estimated from the surface spectrum, i.e. more weathered surfaces have been irradiated for a longer time. The absolute duration of weathering, in turn, tells us the duration of the induced magnetic field that modifies the weathering. The induced magnetic field duration constrains when the interiors were molten, started pluming, and simultaneously induced the magnetic field.

**Key measurements:** A magnetometer will measure the strength and direction of the electric current and Alfvén wing coupling Saturn to Enceladus, as well as measure any induced magnetic field. Plasma spectrometers will measure the beam speeds and energies associated with the coupling. Electric field antennas and magnetic search coils will constrain frequencies and powers of plasma waves that arise as a result of this interaction. Altogether the fields and particles suite will constrain the power of this coupling. Highly sensitive ultraviolet spectrometer will remotely detect the auroral footprints on Saturn's atmosphere, that have only been detected very few times by *Cassini*. In-situ plasma measurements will quantify incident plasma flux and composition irradiated to Enceladus' surface. Ultraviolet, visible, infrared, millimeter/micrometer wave spectra will remotely detect distribution of space weathering activity.

*Summary:* **Saturn is persistently coupled to Enceladus through a large and extensive system of electric currents along Saturn's magnetic field lines. As Enceladus orbits Saturn, it traces a circle of dynamism on Saturn's northern and southern atmospheres that are magnetically conjugate to Enceladus' orbit. The combination of spatial asymmetries on Enceladus' surface and temporal variability of its plume activity means this interaction is highly non-uniform. Enceladus is prone to space weathering by incident plasma, which might be capable of synthesising organic compounds.**

> **Key scientific questions:** What is the strength of Saturn's interaction with Enceladus? In other words, how much energy is transferred between the two bodies? How is Enceladus' surface affected by space weathering? What organic compounds are synthesized by the space weathering process at Enceladus?

# 5. Titan and Enceladus Science in Context

The scientific themes summarised in the previous sections are relevant to a wide range of disciplines within physics, chemistry and biology spanning micro- (e.g. fundamental chemistry) to macro-scales (e.g. evolution). Below is a list of some emerging and fast-growing fields that are relevant to the exploration of Titan and Enceladus.

## Exoplanets and exomoons

The rapid growth of the exoplanetary community is reflected by multiple selected European and international missions, such as *Plato*, *Euclid*, *ARIEL*, *JWST*, etc. The exploration of Titan and Enceladus will uniquely complement these remote observations by providing an in-situ perspective to the knowledge of exoplanetary and/or exomoon composition, structure and formation.

## Ocean worlds and Astrobiology

The topic of ocean worlds in the Solar System has witnessed a 'boom' in the last decade with the selection of ESA's *JUICE* and NASA's *Europa Clipper* missions to the Galilean moons. The potential to understand ocean worlds through the exploration of Titan and Enceladus is limitless. These moons together offer a diverse range of topics in this area that include: (i) Bodies of liquid of various sizes (e.g. lakes on Titan, ocean on Enceladus), (ii) Surface and subsurface bodies, (iii) Depletion and replenishment of bodies (e.g. lakes on Titan), (iv) tidally heated interiors, and (v) Chemical and geological processes (e.g. rock-water interactions). Combined with *JUICE* and *Europa Clipper* findings, the exploration of Titan and Enceladus will bring strong constraints on the presence of liquid water further away from the Sun than previously supposed by the standard habitability zone models in the Solar System, and would provide essential new constraints for the search for habitable worlds outside our Solar System, in exoplanetary systems.

The recent selection of *Dragonfly*, a mission that will send a mobile robotic rotorcraft lander to Titan, for the NASA New Frontiers program, testifies to the interest of the international planetary science community for Saturn's largest moon. Titan is one of the most compelling astrobiology target in the Solar System and *Dragonfly* will assess its prebiotic chemistry and habitability visiting multiple locations at the surface including dunes and a young crater.

**Atmospheres**

The combination of Titan's chemically rich atmosphere and highly variable space environment affords a vast spectrum of atmospheric dynamics, chemistry and cloud formation phenomena that can be explored. In tandem with comparative datasets of other Solar System bodies, a large parameter space can be constructed from which exoplanet and exomoon atmospheres can be characterised.

**Origin and evolution of the Solar System**

The satellites around giant planets can offer clues of how the Solar System evolved in time. For example, the mass-distance relationship of the icy moons suggests a possible linkage to the origin and evolution of giant planets' ring systems (Charnoz et al., 2010). Moreover, the circumstances allowing the capture of small objects into satellites do not exist in the current stage of the Solar System. Understanding the properties of the irregular satellites around giant planets therefore provides a unique window to look into the past of the Solar System evolution.

**Dust in the Solar System and planetary formation**

Dust and gases are the fundamental elements for the formation of stars and planets. Recent studies consider the effects of magnetorotational instability of plasmas. However, the evolution of dusty plasma in the absence of the UV ionization at the center of the protoplanetary disk and the interactions between dust and plasma are missing links. It is postulated that dust-plasma interaction must influence the nucleation of grains as well as their subsequent agglomeration. The Enceladus plume and Titan's atmosphere are the sites to investigate the nature of charged dust and its interaction with plasma, and may give hints to addressing questions on planetary formation mechanisms.

# 6. Titan and Enceladus Missions

**Mission Considerations**

*Payload*

The diverse scientific opportunities highlighted in Sections 3 and 4 call for a range of instruments, most of which can participate in more than one experiment. In-situ instrumentation is required for direct sampling of Titan's atmosphere and lakes, Enceladus' plume and their interaction with Saturn's environment. Their capabilities include, but are not limited to: Mass spectroscopy of ions and neutrals, plasma analysis (ions, electrons), aerosol and dust detection, electric and magnetic field direction, frequency and power. Multi-spectral remote sensing instrumentation is required for the characterisation of Titan's atmosphere, their surfaces and interiors. These capabilities include, but are not limited to: Radar imaging and sounder, ultraviolet, visible, infrared, millimeter and micrometer spectroscopy and imaging, gravity radiometry, seismometer, thermal sensing. These instruments have strong European heritage from previous, existing and upcoming missions such as *JUICE*, *Cassini/Huygens*, *Rosetta*, *BepiColombo*, etc.

*Planetary protection*

Titan and Enceladus hold the likelihood of hosting biosignatures, thus making planetary protection considerations necessary. Fortunately, collecting samples from Enceladus' plume means possible biosignatures originating from the moon's interior can be obtained without penetrating the surface. This greatly mitigates risks associated with planetary protection, as well as mission complexities and costs in general. For this reason, we argue that Enceladus poses the least risk for the search for biosignatures in the outer Solar System.

*Radiation*

Saturn has relatively weak radiation belts, thus making radiation considerations manageable. The outermost edge of the main radiation belt is situated at 3.5 Saturn radii, which is planetward of Enceladus' orbit (4 Saturn radii) and Titan's orbit (20 Saturn radii).

## Architectures

### L-class or M-class (Titan and Enceladus Orbiter; Bioinspiration and Biomimetrics)

An orbiter with a probe/lander would cover all the identified scientific goals. This will ensure the spatial coverage required to fully map the surfaces of the satellites and characterise their interiors using gravity and induction measurements. A Titan orbiter would also serve as a communication link between Earth and a probe/lander on the ground or lake. Titan has a relatively large mass and is far enough away from Saturn to impose a reasonable $\Delta V$ cost. Enceladus, on the other hand, possesses a very low mass (0.8% Titan's mass) and orbits deep within Saturn's gravity well thus demanding an orbit insertion $\Delta V$ that is prohibitively large. Efficient tour designs have been explored such as a leveraging tour with Titan, Rhea, Dione and Tethys to reach Enceladus orbit. This was found to require less than half of the $\Delta V$ of a direct Titan-Enceladus transfer (Strange et al., 2009). Free-return cycler trajectories are also possible, where a spacecraft shuttles between Enceladus and Titan using little or no fuel (Russell and Strange, 2009).

Numerous diverse Titan missions have been proposed in the last decade (Reh, 2009; Oleson et al., 2015; Barnes et al., 2012). A mission to the Titan-Enceladus system, *TandEM*, has been extensively studied as an L-class mission by Coustenis et al. (2009), where they explored the possibilities of a hot air balloon (Titan), mini-probes (Titan) and penetrators (Titan and Enceladus). In June 2019, *Dragonfly* was selected as NASA's next New Frontiers mission. The mission will send a robotic rotorcraft lander to Titan in order to explore prebiotic chemistry and habitability. Such heavier-than-air flight is made possible by Titan's thicker atmosphere (1.5x that of the Earth's) and smaller gravitational acceleration.

A recent concept of a versatile aerial-aquatic robotic aircraft provides the capability of in-situ near-surface atmosphere and surface liquid observations (McKevitt, 2019). The operation has heritage in robotic work inspired by observations of the natural world – the field of bioinspiration and biomimetrics. A 'plunge diving' manoeuvre, inspired by the gannet seabird (Liang et al., 2013), involves the aircraft plunging nose-first into the surface of Titan's lake. The vehicle is capable of relaunching and ejecting a mass of liquid collected from the area of launch, as shown in Figure 9 (Siddall and Kovac, 2014). Through this, measurements of the compositions of Titan's lakes and near-surface atmosphere can be achieved. Entry and descent data can also be used to perform upper and mid-atmospheric observations, in a similar way to the *Huygens* descent.

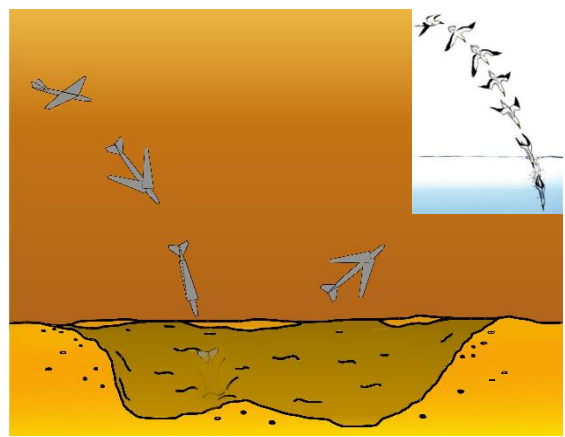

**Figure 9** – Impression of a 'plunge diving' manoeuvre by an aerial-aquatic aircraft inspired by the gannet seabird (inset). Inset adapted from Liang et al. (2013).

A Saturn orbiter with multiple satellite flybys would also provide significant scientific return on the aforementioned themes. The Saturn orbiter would have state-of-the-art payload that exclusively affords satellite science.

### S-class or F-class (Plume flyby)

A flyby mission could provide significant scientific return, however this would focus on a single theme, namely habitability. A spacecraft with focused payload would fly through Enceladus' plume and conduct ultra-high resolution and sensitivity measurements of its composition in search for biosignatures.

## Enabling technology

### Radioactive power sources

Electrical power sources for outer planet missions are a key enabling technology. Electrical power and spacecraft heating are issues for any mission beyond the orbit of Jupiter due to the low solar irradiance at such distances. This is further complicated by the decrease in solar panel efficiency at low intensities (LILT). To supply sufficient, long-term, and uninterrupted power, solar panels of an unfeasibly large area would be required. This makes both Radioisotope Thermoelectric Generators (RTGs) and Radioisotope Heating Units (RHUs) necessary to power and heat the spacecraft, respectively. Within Europe, $^{241}$AM is favoured over $^{238}$Pu as a source of radioisotopes. Despite $^{241}$AM having a lower power density that that of $^{238}$Pu, its longer half-life and more cost-effective production makes it the economical alternative. A mission to explore Titan and Enceladus would greatly benefit from an independent European power source.

### Electromagnetic cleanliness

EMC issues should be explored to satisfy requirements imposed by some of the listed payload elements on the spacecraft, e.g. plasma packages and magnetometers. This would be especially crucial to resolve any induced and fossil magnetic fields that may be found on Titan.

### Telecommunications

Given the range of the proposed target, science data return will be limited by bandwidth. Most deep space missions use X-band links, while few also use Ka-band, to transmit telemetry. Nowadays, NASA's Deep Space Network (DSN) 70-m radio antennas provide the maximum rates. Since *Cassini* relied on NASA's DSN, a mission to Titan and Enceladus would therefore require similar capabilities to achieve the minimum science data return. Further studies of new or upgraded telecommunications technologies are welcomed. This includes both Earth-direct and intra-spacecraft (i.e. relay between probe and spacecraft) communications.

### Autonomous guidance, navigation and control

Autonomous GN&C systems are required whenever position and attitude must be known precisely and updated quickly. Proximity missions, particularly small body proximity, will require onboard autonomous GN&C to detect and avoid surface hazards and especially to minimise planetary protection risks. This technology more broadly applies to flyby, small body rendezvous and orbiting, landing, atmospheric entry and "touch and go" sampling, for example to cope with severe and unpredictable contact forces and torques.

### Mass spectrometry and dust analysis

The surprise discovery of Enceladus' plume by Cassini meant the mass spectrometer and dust analyser onboard were not specifically designed for such measurements. The composition of the plume was determined using a low resolution mass spectrometer. A more detailed analysis of the plume in search for complex biosignatures would require a mass spectrometer with higher capability of measuring masses up to 1000 u, with a high resolution exceeding 24,000 m/$\Delta$m and a high sensitivity of one part per trillion. The state-of-the-art mass spectrometer should not only be capable of identifying complex organic chains, but also differentiating their chirality. Similarly, the dust and ice analysers should have higher capabilities, resolution and sensitivity to pick up individual ice and dust grains with micron and sub-micron diameters. These requirements have been studied in detail by the Enceladus Life Finder team [Lunine et al., 2015].

## 7. Summary and Perspectives

This white paper briefly describes outstanding questions pertaining to Titan and Enceladus - legacies of the successful *Cassini-Huygens* mission. We make the case that such questions are not merely specific to these two mysterious systems but have much broader and deeper implications for humakind's outstanding questions at large of habitability in the Solar System. For these reasons, **we recommend the acknowledgement of Titan and Enceladus as priorities for ESA's Voayage 2050 programme** and to combine efforts, in science and technology, with international agencies to launch a dedicated mission to either or both tagets, much like ESA's key involvement in some of the most successful planetary missions like *Cassini-Huygens*.

*The core proposers fully support complementary white papers led by S. Rodriguez and G. Choblet advocating Titan and Enceladus science, respectively. Altogether, these white papers are a testament to the size, strength and diversity of the Titan and Enceladus community.*

# Core Proposing Team

**Ali Sulaiman**
University of Iowa, USA.

**Nicholas Achilleos**
University College London, UK.

**Sushil Atreya**
University of Michigan, USA.

**Cesar Bertucci**
Universidad de Buenos Aires, Argentina.

**Andrew Coates**
Mullard Space Science Laboratory, UK.

**Michele Dougherty**
Imperial College London, UK.

**Lina Hadid**
Laboratoire de Physique des Plasmas,
France.

**Candice Hansen**
Planetary Science Institute, USA.

**Mika Holmberg**
Institut de Recherche en Astrophysique et
Planétologie, Toulouse, France.

**Hsiang-Wen (Sean) Hsu**
University of Colorado, Boulder, USA.

**Tomoki Kimura**
Tohoku University, Japan.

**William Kurth**
University of Iowa, USA.

**Alice Le Gall**
LATMOS, Université Versailles Saint
Quentin, France.

**James McKevitt**
Imperial College London, UK.

**Michiko Morooka**
Swedish Institute of Space Physics
Uppsala, Sweden.

**Go Murakami**
Japan Aerospace Exploration Agency,
Japan.

**Leonardo Regoli**
University of Michigan, USA.

**Elias Roussos**
Max Planck Institute for
Solar System Research, Germany.

**Joachim Saur**
University of Cologne, Germany

**Oleg Shebanits**
Imperial College London, UK.

**Anezina Solomonidou**
LESIA, Observatoire de Paris, France

**Jan-Erik Wahlund**
Swedish Institute of Space Physics
Uppsala, Sweden.

**J. Hunter Waite**
Southwest Research Institute, USA.